\documentclass{article}
\usepackage{graphicx}
\usepackage{amsmath}
\usepackage{amssymb}
\usepackage{bm}
\usepackage{upgreek}
\usepackage{booktabs}
\usepackage{tabularx}
\usepackage[colorlinks=false]{hyperref}
\usepackage{url}
\usepackage{footmisc}
\usepackage{authblk}
\usepackage{nicefrac}
\usepackage{tikz}
\usepackage{pgffor}
\usetikzlibrary{positioning}
\usetikzlibrary{calc}
\usetikzlibrary{intersections}
\usepackage[labelfont=bf]{caption}

\usepackage[left=2.5cm, right=2.5cm]{geometry}

\usepackage[defernumbers=true,style=nature,backend=bibtex]{biblatex}
\makeatletter
\def\dif{\@ifnextchar[{\@with}{\@without}}
\def\@with[#1]#2{
  \ensuremath{\frac{\foreach \x in {#2}{\mathrm{d}\x\,}}{\foreach \x in {#1}{\mathrm{d}\x\,}}}
}
\def\@without#1{
  \ensuremath{%
    \ifx\hfuzz#1\hfuzz
    \mathrm{d}
    \else
    \foreach \x in {#1}{\mathrm{d}\x\,}
    \fi
    }
}
\makeatother

\newcommand{\hess}{H.E.S.S.}
\newcommand{\sd}{\ensuremath{_\textsc{sd}}}
\newcommand{\ts}{\ensuremath{_\textsc{ts}}}
\newcommand{\be}{\begin{equation}}
\newcommand{\ee}{\end{equation}}
\newcommand{\ve}{\ensuremath{\varepsilon}}
\newcommand{\inj}{\ensuremath{_\textsc{inj}}}
\newcommand{\cut}{\ensuremath{_\textsc{cut}}}

\newcommand{\ad}{\ensuremath{_\textsc{ad}}}
\newcommand{\ic}{\ensuremath{_\textsc{ic}}}
\newcommand{\ph}{\ensuremath{_\textsc{ph}}}
\newcommand{\fir}{\ensuremath{_\textsc{fir}}}
\newcommand{\nir}{\ensuremath{_\textsc{nir}}}
\newcommand{\cmb}{\ensuremath{_\textsc{cmbr}}}
\newcommand{\ssc}{\ensuremath{_\textsc{ssc}}}
\newcommand{\syn}{\ensuremath{_\textsc{syn}}}

\newcommand{\sigmat}{\ensuremath{\sigma_\textsc{t}}}
\newcommand{\obs}{\ensuremath{_\textsc{obs}}}

\newcommand{\ort}[1]{\ensuremath{\bm{\hat #1}}}
\newcommand{\bp}{\ensuremath{\bm{p}}}
\newcommand{\br}{\ensuremath{\bm{r}}}

\newcommand{\bb}{\ensuremath{\bm{\beta}}}
\newcommand{\upi}{\uppi}
\newcommand{\Db}{\ensuremath{{\cal D}}}

\newcommand{\G}{\ensuremath{\gamma}}

\begin{document}

\title{Resolving the Crab pulsar wind nebula at teraelectronvolt energies}

\author[1]{\tiny H.E.S.S. Collaboration: H.~Abdalla}
\author[2,3,4]{\tiny F.~Aharonian}
\author[2]{\tiny F.~Ait~Benkhali}
\author[5]{\tiny E.O.~Ang\"uner}
\author[6]{\tiny M.~Arakawa}
\author[1]{\tiny C.~Arcaro}
\author[7]{\tiny C.~Armand}
\author[8,1]{\tiny M.~Backes}
\author[1]{\tiny M.~Barnard}
\author[9]{\tiny Y.~Becherini}
% ...
\affil[1]{Centre for Space Research, North-West University, Potchefstroom 2520, South Africa }
\affil[2]{Max-Planck-Institut f\"ur Kernphysik, P.O. Box 103980, D 69029 Heidelberg, Germany }
\affil[3]{Dublin Institute for Advanced Studies, 31 Fitzwilliam Place, Dublin 2, Ireland }
\affil[4]{National Academy of Sciences of the Republic of Armenia,  Marshall Baghramian Avenue, 24, 0019 Yerevan, Republic of Armenia  }
\affil[5]{Aix Marseille Universit\'e, CNRS/IN2P3, CPPM, Marseille, France }
\affil[6]{Department of Physics, Rikkyo University, 3-34-1 Nishi-Ikebukuro, Toshima-ku, Tokyo 171-8501, Japan }
\affil[7]{Laboratoire d'Annecy de Physique des Particules, Univ. Grenoble Alpes, Univ. Savoie Mont Blanc, CNRS, LAPP, 74000 Annecy, France }
\affil[8]{University of Namibia, Department of Physics, Private Bag 13301, Windhoek, Namibia, 12010 }
\affil[9]{Department of Physics and Electrical Engineering, Linnaeus University,  351 95 V\"axj\"o, Sweden }
% ...

\maketitle

\begin{refsection}
  
\textbf{
  The Crab nebula is one of the most studied cosmic particle
  accelerators, shining brightly across the entire electromagnetic
  spectrum up to very high-energy gamma
  rays~\autocite{2008ARA&A..46..127H,2014RPPh...77f6901B}. It is known
  from radio to gamma-ray observations that the nebula is powered by a
  pulsar, which converts most of its rotational energy losses into a
  highly relativistic outflow. This outflow powers a pulsar wind
  nebula (PWN), a region of up to 10~light-years across, filled with
  relativistic electrons and positrons. These particles emit
  synchrotron photons in the ambient magnetic field and produce very
  high-energy gamma rays by Compton up-scattering of ambient
  low-energy photons. While the synchrotron morphology of the 
  nebula is well established, it was up to now not known in which
  region the very high-energy gamma rays are
  emitted~\autocite{1989ApJ...342..379W,2000A&A...361.1073A,HessCrab,2008ApJ...674.1037A,KevinMeagherfortheVERITAS:2015vya,2017ApJ...843...39A}.
  Here we report that the Crab nebula has an angular extension at
  gamma-ray energies of 52 arcseconds (assuming a Gaussian source
  width), significantly larger than at X-ray energies. This result
  closes a gap in the multi-wavelength coverage of the nebula,
  revealing the emission region of the highest energy gamma
  rays. These gamma rays are a new probe of a previously inaccessible
  electron and positron energy range. We find that simulations of the
  electromagnetic emission reproduce our new measurement, providing a
  non-trivial test of our understanding of particle acceleration in
  the Crab nebula.
}

The Crab pulsar's relativistic outflow is a cold, non-turbulent wind of charged
particle pairs, electrons and positrons (commonly called electrons in
the following). At a distance of 0.5~light-years from
the pulsar, the wind is heated up by passing through a
\emph{termination shock}. Beyond this point, most of the electrons in
the shocked wind have energies of 100--300\,GeV (gigaelectronvolts,
$10^{9}$\,eV), with maximum energies reaching up to PeV
(petaelectronvolts, $10^{15}$\,eV) energies. For a black body, such
average thermal particle energies would be reached at extreme
temperatures of $10^{15}$\,K, clearly demonstrating that this system
is a non-thermal particle accelerator. The shocked wind continues
propagating away from the pulsar, reaching at the present epoch out to
distances of several light-years.

\begin{figure*}
  \centering
  \includegraphics[width=\textwidth]{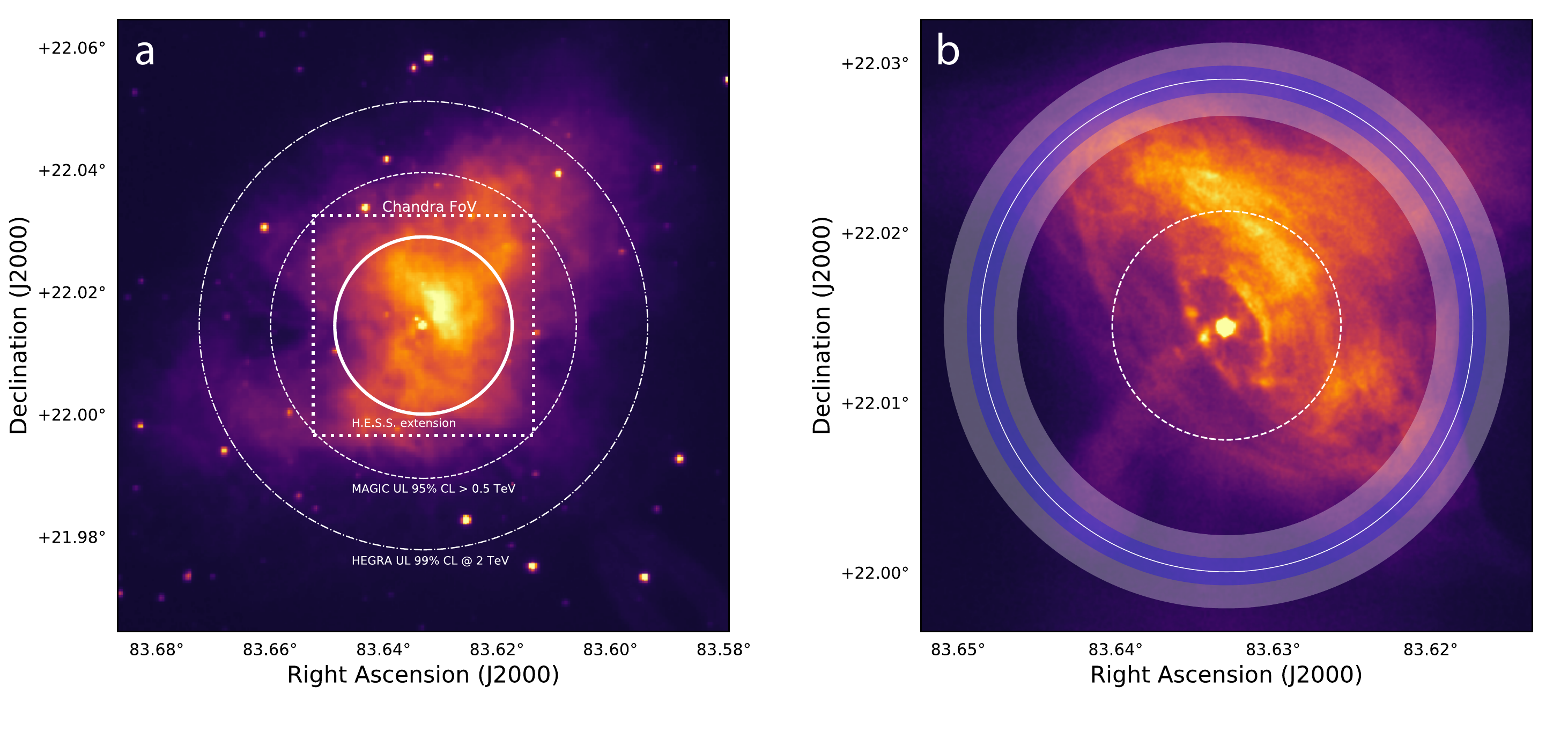}
  \caption{\textbf{Images of the Crab nebula.} \textbf{a:} UV
    ($\lambda = 291\,$nm) image recorded with the Optical-UV Monitor
    onboard \emph{XMM-Newton}~\autocite{2017_Dubner}. The MAGIC and
    HEGRA extension upper limits of $2.2'$~\autocite{2008ApJ...674.1037A}
    and $1.5'$~\autocite{2000A&A...361.1073A} are drawn as dash-dotted
    and dashed lines, respectively. The extent of the sky region shown
    in \textbf{b} is indicated as dotted square, and the
    \hess\ extension (two-dimensional Gaussian $\sigma$ corresponding
    to 39\% of the measured events) is drawn as a solid circle. All
    circles are centred on the Crab pulsar position for illustration
    purposes, in the fit procedure determining the \hess\ extension
    described in the main text the centroid position is left
    free. \textbf{b:} \emph{Chandra} X-ray
    image~\autocite{2000ApJ...536L..81W} (courtesy of M.~C.~Weisskopf
    and J.~J.~Kolodziejczak). The \hess\ extension is shown as solid
    white circle overlaid on top of shaded annuli indicating the
    statistical and systematic uncertainties of our measurement. The
    \emph{Chandra} extension, corresponding to 39\% of the X-ray
    photons, is given as dashed white circle.}
  \label{fig:pwn}
\end{figure*}

This part of the wind behind the termination shock is a region filled
with radiating electrons producing radio to gamma-ray emission and is
known as the Crab pulsar wind nebula (PWN, shown in
Fig.~\ref{fig:pwn}).  While the radio synchrotron emission of the Crab
PWN was discovered in the 1960's~(see
e.g. ref.~\autocite{2014RPPh...77f6901B} for a recent review), the
Inverse Compton (IC) component of the PWN at GeV to TeV
(teraelectronvolts, $10^{12}$\,eV) photon energies was only discovered
in 1989 with the Whipple
telescope~\autocite{1989ApJ...342..379W}. Today, the Crab nebula is
the brightest steady source of TeV gamma rays in the sky and is the
standard candle of gamma-ray astronomy; it is regularly observed by
all gamma-ray
telescopes~\autocite{2010ApJ...708.1254A,2014A&A...562L...4H,2015JHEAp...5...30A,2017APh....91...34A,2017ApJ...843...39A}.

The spectrum and morphology of the PWN depend on the structure of the
post-shock magnetohydrodynamic (MHD) flow. Observations in the X-ray
energy band with the \emph{Chandra} X-ray
Observatory~(\emph{Chandra})~\autocite{2000ApJ...536L..81W} revealed a
complex structure consisting of a bright torus and a narrow jet
emerging in the direction perpendicular to the torus plane~(see
Fig.~\ref{fig:pwn}, right). The X-ray structure suggests that the
underlying MHD flow is nearly axisymmetric. The apparent deviation
from this symmetry in the X-ray image is mostly due to Doppler
boosting, which enhances the X-ray emission of fluid elements moving
towards us. MHD instabilities developing in the post-shock region also
alter the symmetry, but at a less important level.

In the model we use here (see ref.~\autocite{2002MNRAS.336L..53B} and
the \emph{Supplementary Information}), which is based on the seminal
work of Kennel and
Coroniti~\autocite{1984ApJ...283..694K,1984ApJ...283..710K}, the key
observational features are reproduced by approximating the pulsar wind
by a two-dimensional axisymmetric MHD flow propagating into a limited
solid angle close to the torus plane. We combine this with a
three-dimensional treatment of the radiation throughout the nebula to
account for the Doppler boosting and orientation of the magnetic
field. To this end, a high-energy particle distribution following a
power law in energy forms at the termination shock. These particles,
confined in fluid elements, are advected through the nebula by the MHD
flow. Accounting for the evolution of the magnetic field, various
target photon fields, and the changing rate of adiabatic cooling, the
energy distribution of high-energy particles in each point of the
nebula is computed. Taking into account local magnetic and photon
fields as well as the Doppler boosting, this allows us to derive the
surface brightness, which can then be directly compared with
observations in different energy bands. We note that in the model the
IC emission is dominated by upscattering of photons of the cosmic
microwave background and the entire spectrum of synchrotron photons
emitted by the same population of high-energy electrons (with two
electron populations, the \emph{wind} and the relic \emph{radio}
electrons, as detailled in ref.~\autocite{1996MNRAS.278..525A}).

\begin{figure*}
  \centering
  \includegraphics[width=0.9\textwidth]{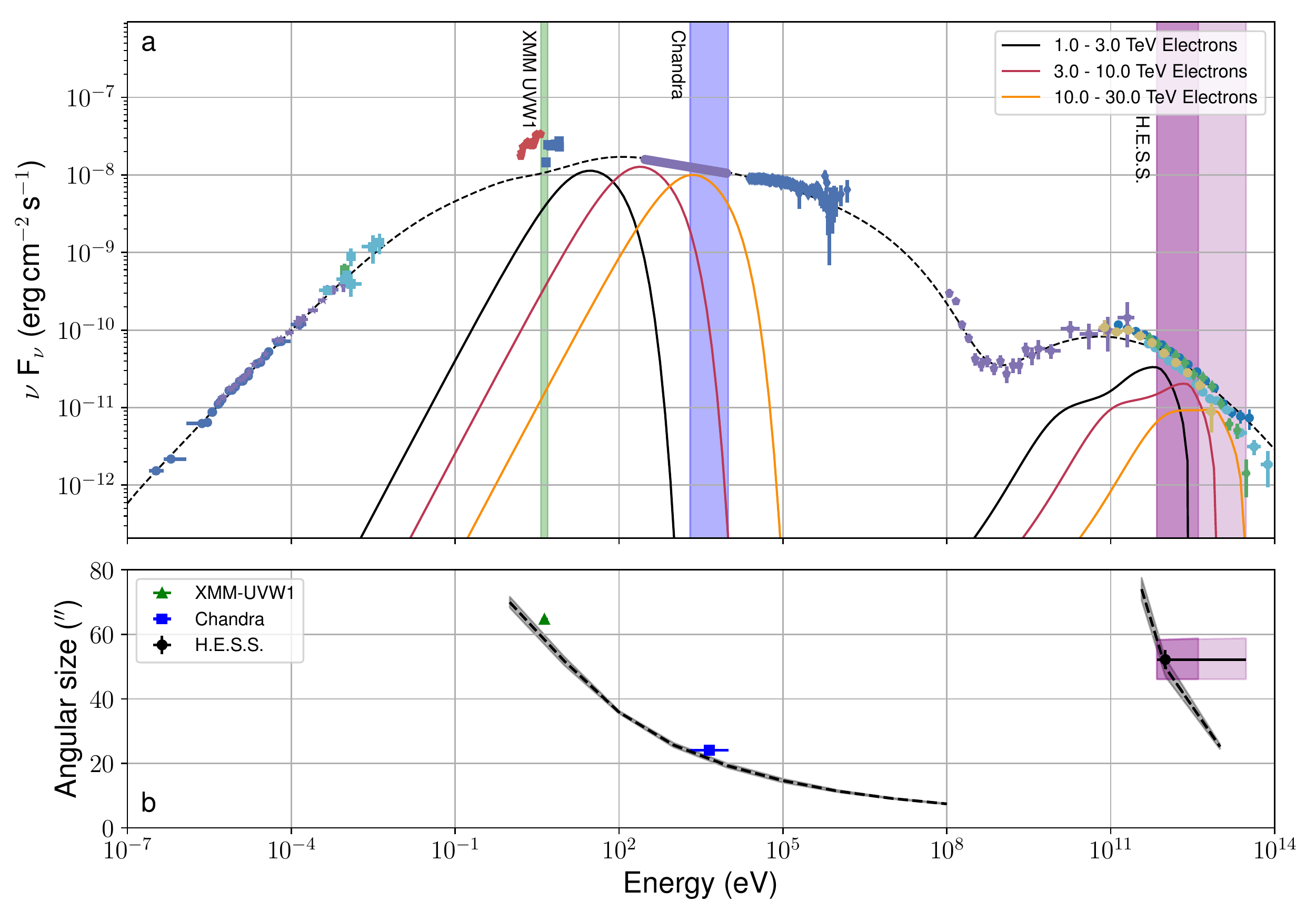}
  \caption{\textbf{Spectral energy distribution (SED) along with the
      measured and predicted extensions of the Crab pulsar wind
      nebula.} \textbf{a:} The SED is shown as dashed line. To
    illustrate the contribution of electrons of different energies to
    the radiation, the coloured lines show the synchrotron and IC
    radiation for electrons in the energy bands 1--3~TeV (black),
    3--10~TeV (red), and 10--30~TeV (yellow). The vertical bands
    indicate the measurement ranges of instruments in the UV (green),
    X-ray (blue), and TeV gamma-ray regime (purple). The dark purple
    part of the \hess\ band indicates the energy range covered by 90\%
    of the measured gamma-ray photons. The range of the remaining 10\%
    of the highest energy photons is given as the light purple
    band. The data points from low to high energies are taken from
    refs.~\autocite{2010ApJ...711..417M,1986A&A...167..145M,2002A&A...386.1044B,1993A&A...270..370V,1992ApJ...395L..13H,1981ApJ...245..581W,2005SPIE.5898...22K,2009ApJ...704...17J,2012ApJ...749...26B,2004ApJ...614..897A,HessCrab,2008ApJ...674.1037A,KevinMeagherfortheVERITAS:2015vya}. Note
    that in the optical domain, the data points are above the SED
    indicative of a substantial contribution from thermal
    emission. \textbf{b:} The predicted (dashed line and grey shaded
    area, corresponding to the uncertainty) and measured (markers)
    extensions are plotted for various photon energies. The predicted
    extensions are the best-fit values of our model to the
    \emph{Chandra} and \hess\ data; the grey shaded uncertainty band
    results from up and down variations of 1 standard deviation of the
    fit parameters. The measured UV and X-ray extensions are
    determined by convolving the respective PWN images with the \hess\
    PSF and applying the same likelihood fit procedure described in
    the main text. The purple boxes indicate the \hess\ energy range,
    their vertical size corresponds to the systematic uncertainty. All
    error bars are 1 standard deviation.}
  \label{fig:sed}
\end{figure*}

In our model, the structure of the nebula depends on three parameters:
the radius of the termination shock, the magnetisation (defined as the
ratio of electromagnetic to particle energy flux) of the wind at the
termination shock, and the wind opening angle defined as the solid
angle into which the bulk of the energy is ejected close to the torus
plane. We note that there are other physical parameters of the system,
like the spin-down luminosity of the pulsar, whose measured values are
used as input to our model. With a suitable choice of parameters, the
model reproduces the broadband spectral energy distribution (SED)
well, as shown in Fig.~\ref{fig:sed}. The model also predicts that the
size of the nebula varies strongly with the energy of the emitting
electrons. Since higher energy electrons suffer more severe radiation
losses from both synchrotron radiation and IC scattering processes,
they propagate shorter distances before they lose energy via
radiation.  This energy dependence is seen by current X-ray telescopes
in the synchrotron radiation of the PWN up to
40~keV~\autocite{2015ApJ...801...66M}, and is also clearly seen in the
data shown in Fig.~\ref{fig:sed},~bottom. In the gamma-ray IC
radiation domain, the PWN emission up to now appeared point-like and
only upper limits on the size could be
derived~\autocite{2000A&A...361.1073A,HessCrab,2008ApJ...674.1037A} or
evidence be claimed at GeV energies~\autocite{2018ApJS..237...32A}.

With the new High Energy Stereoscopic System~(\hess) measurement shown
in Fig.~\ref{fig:pwn}, we establish the IC PWN extension, at photon
energies eight orders of magnitude above the previously highest energy
morphology measurement of the Crab PWN in X-rays. This measurement of
the IC extension of the Crab PWN provides a stringent test of our
understanding of high-energy particle propagation and radiation
models. Using 22\,hours of observations collected over 6\,years, we
employ advanced analysis techniques to reconstruct the gamma-ray image
of the Crab PWN. We compare this image to that of a simulated
gamma-ray source, taking for each observation the exact hardware
status of the \hess\ telescopes and all observation conditions into
account in the simulations. Simulating the exact state of the
telescope system at any given time was not previously done, but
allowed us here to increase the precision to the level needed to
discover the extension of the Crab PWN at TeV gamma-ray energies
(further details are given in \emph{Methods}).

\begin{figure*}
  \centering
  \includegraphics[width=0.95\textwidth]{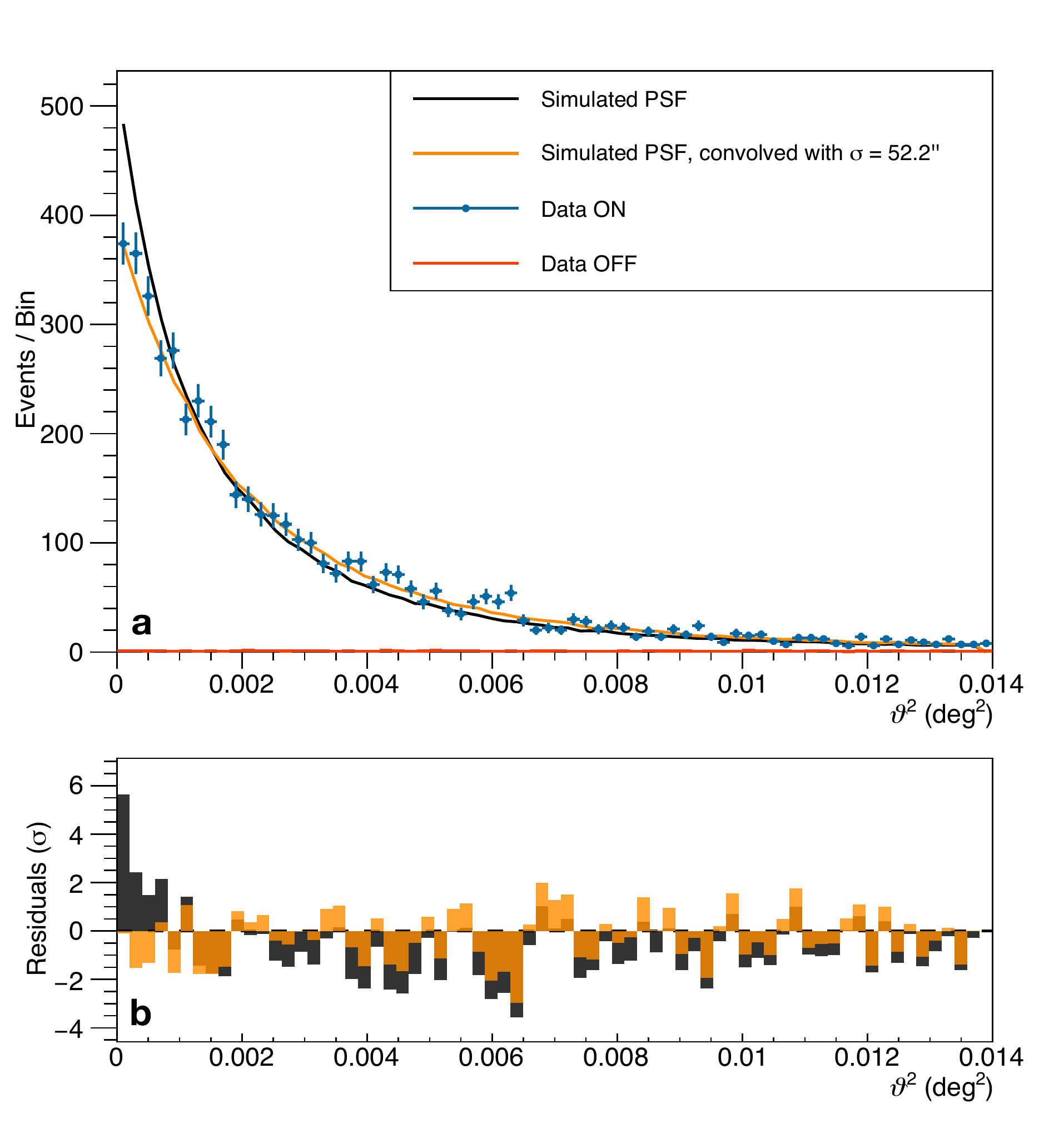}
  \caption{\textbf{a:} Histogram of reconstructed directions of gamma
    rays from the Crab nebula (\textit{Data ON}, blue). The estimated
    background determined in empty regions of the sky is also shown
    (\textit{Data OFF}, red). For comparison, the simulated angular
    resolution function (\textit{point spread function, PSF}, black)
    for this dataset as well as the function convolved with the
    best-fit Gaussian (yellow) are shown. The error bars are 1
    standard deviation.\textbf{b:} Significance of the bin-wise
    deviation (\textit{Data - MC}) of the data when compared to the
    PSF (black) and the convolved one (orange).}
  \label{fig:ThetaSq}
\end{figure*}

The comparison of data to simulations reveals that the nebula is
extended~(see Fig.~\ref{fig:ThetaSq}). We can reproduce the data only
by simulating an extended source. We do this by convolving the angular
resolution function with a two-dimensional Gaussian with a best-fit
value of $\sigma_{2\mathrm{D,G}} = 52.2'' \pm 2.9''_{\mathrm{stat}}
\pm 6.6''_{\mathrm{sys}}$. The systematic uncertainty is related to
the calibration and analysis method, to the spectral shape used to
simulate the angular resolution, and to the fit method. The resulting
radial distribution of gamma rays compared to the simulated angular
resolution function, as well as the resolution function convolved with
a Gaussian, is shown in Fig.~\ref{fig:ThetaSq}. The event distribution
is clearly incompatible with a point-like source, while a convolution of the
resolution function with a Gaussian drastically improves the fit to
the data by 9~standard deviations. We therefore conclude that we
measure the Crab nebula as a substantially extended gamma-ray source
at photon energies above 700~GeV.

The extension we measure is smaller than that seen in ultraviolet~(UV)
light and significantly larger than that seen in hard X-rays (see
Fig.~\ref{fig:pwn}). This can be understood by considering the
energies of the electrons producing the synchrotron and IC emission,
respectively. As shown in Fig.~\ref{fig:sed}, lower energy electrons
are emitting the UV synchrotron photons, medium energy electrons are
emitting the IC gamma rays, and higher energy electrons are
responsible for the synchrotron X-ray emission measured in the
\emph{Chandra} image. Thus, the difference in size in the different
wavebands is compatible with the energy dependent radiation losses of
the parent electrons discussed above: higher energy electrons
propagate shorter distances than lower energy ones~(see
Fig.~\ref{fig:sed},~bottom).

By splitting our data into two parts, below and above 5~TeV, we have
also searched for energy dependent changes of the TeV gamma-ray
extension. Such changes are ultimately also expected to show up in TeV
gamma-ray data. While the Crab PWN is significantly extended in both
energy bins, our data are currently not precise enough to establish
this energy dependence of the PWN extension.

As Fig.~\ref{fig:sed}~(top) shows, the PWN morphology reflecting an
electron energy range of 1--10~TeV is now probed for the first time
with the new \hess\ image. The majority of the photons measured (dark
purple vertical band) probes this electron energy range, which is
inaccessible via 100~eV synchrotron emission due to absorption by
interstellar matter. The measurement of the TeV gamma-ray morphology
of the Crab PWN is therefore the only way to trace such 1--10~TeV
electrons.

While the simplest one-dimensional MHD
simulations~\autocite{1984ApJ...283..694K,1984ApJ...283..710K,1996MNRAS.278..525A,2002MNRAS.336L..53B}
are known to reproduce the basic characteristics of the Crab PWN, we
have now verified that with the model we
use~\autocite{2002MNRAS.336L..53B} we can consistently reproduce the
measured synchrotron and IC morphology of the PWN including our new
measurement~(see Fig.~\ref{fig:sed},~bottom). We find a best-fit model
for the wind solid angle of \(\sim6.5\) steradians and a termination
shock radius of \(\sim0.13\)~parsec, both in good agreement with the
\emph{Chandra} X-ray
data~\autocite{2000ApJ...536L..81W,2012ApJ...746...41W}. The
magnetisation of the wind is found to be \(\sim0.5\)\%, and is thus at
the same level as previously found to reproduce the SED and
synchrotron
morphology~\autocite{1996MNRAS.278..525A,2010A&A...523A...2M}. We note
that the magnetisation value is affected by the assumed radial flow
velocity and a lack of turbulence in our approach and can therefore
not be directly compared to higher dimensional MHD
simulations~\autocite{2004AA...421.1063D,2004MNRAS.349..779K,2006MNRAS.368.1717B,2006A&A...453..621D,2008A&A...485..337V,2013MNRAS.431L..48P}
(see also the \emph{Supplementary Information}). 

With this measurement we establish the extension of the Crab nebula at
TeV gamma-ray energies and provide a new probe of the distribution of
1--10~TeV electrons. This closes a gap in the multi-wavelength
coverage of this icon of high-energy astrophysics and provides a
non-trivial test of our understanding of particle propagation and
photon emission at very high energies.

\printbibliography[resetnumbers=false]
  
\textbf{Acknowledgements:} The support of the Namibian authorities and
of the University of Namibia in facilitating the construction and
operation of H.E.S.S. is gratefully acknowledged, as is the support by
the German Ministry for Education and Research (BMBF), the Max Planck
Society, the German Research Foundation (DFG), the Helmholtz
Association, the Alexander von Humboldt Foundation, the French
Ministry of Higher Education, Research and Innovation, the Centre
National de la Recherche Scientifique (CNRS/IN2P3 and CNRS/INSU), the
Commissariat \`{a} l'\'{e}nergie atomique et aux \'{e}nergies alternatives (CEA),
the U.K. Science and Technology Facilities Council (STFC), the Knut
and Alice Wallenberg Foundation, the National Science Centre, Poland
grant no. 2016/22/M/ST9/00382, the South African Department of Science
and Technology and National Research Foundation, the University of
Namibia, the National Commission on Research, Science \& Technology of
Namibia (NCRST), the Austrian Federal Ministry of Education, Science
and Research and the Austrian Science Fund (FWF), the Australian
Research Council (ARC), the Japan Society for the Promotion of Science
and by the University of Amsterdam. We appreciate the excellent work
of the technical support staff in Berlin, Zeuthen, Heidelberg,
Palaiseau, Paris, Saclay, Tübingen and in Namibia in the construction
and operation of the equipment. This work benefited from services
provided by the H.E.S.S. Virtual Organisation, supported by the
national resource providers of the EGI Federation.

\textbf{Author Contributions:} D.~Berge, J.~Hahn, M.~Holler,
D.~Khangulyan, D.~Parsons have analysed and interpreted the data, and
prepared the manuscript. The whole H.E.S.S. collaboration has
contributed to the publication with involvement at various stages
ranging from the design, construction and operation of the instrument,
to the development and maintenance of all software for data handling,
data reduction and data analysis. All authors have reviewed,
discussed, and commented on the present results and on the manuscript. 

\textbf{Author Information:} Author Information Reprints and
permissions information is available at \url{www.nature.com/reprints}. The
authors declare no competing financial interests. Correspondence and
requests for materials should be addressed to the \hess\ Collaboration
(\href{mailto:contact.hess@hess-experiment.eu}{contact.hess@hess-experiment.eu}). 

\clearpage

\end{refsection}

\clearpage

\begin{refsection}
  
\section*{Methods} The dataset used here was recorded with the High
Energy Stereoscopic System (H.E.S.S.) array of telescopes. H.E.S.S. is
an array of five imaging atmospheric Cherenkov telescopes. Such
telescopes reconstruct cosmic gamma rays by recording images of
Cherenkov light of the air showers that develop when a cosmic gamma
ray smashes into the atmosphere. Such air showers are cascades of
secondary charged particles, mostly electrons and positrons, which are
created when gamma rays penetrate the atmosphere.  The charged
particles emit Cherenkov light, which in turn can be used to
reconstruct the direction and energy of the primary gamma ray with
telescopes like the H.E.S.S. array. The system consists of four
telescopes with 108\,m$^2$ mirror area and 15\,m focal length and a
single 614\,m$^2$ telescope of 36\,m focal length. \hess\ is situated
in the Khomas highlands of Namibia and is in the five-telescope
configuration for observations near zenith sensitive to gamma-ray
photons in the energy range from around 50\,GeV to around 50\,TeV. The
analysis presented here uses only data from the four small telescopes
(which have a larger energy threshold of 100~GeV near zenith). As the
Crab nebula is such an important gamma-ray source it is regularly
monitored by H.E.S.S. From this large monitoring dataset recorded over
the course of 10 years, 22 hours of observations fulfil tight quality
selection criteria aimed at optimising the angular resolution of the
system and are used in this study (see Supplementary Table~1).

The data were analysed with the analysis technique introduced in
ref.~\autocite{2009_deNaurois}. This method is based on a
semi-analytical air-shower model, which is fit to the recorded
air-shower images to yield the primary gamma-ray direction and
energy. To improve the angular resolution of the standard analysis
configuration, only well reconstructed gamma-ray candidates are
considered further.

The analysis was conducted in three ranges in reconstructed energy,
once using all events reconstructed between 0.7~and~30~TeV, and once
in two separate energy bins from 0.7~to~5~TeV and from 5~to~30~TeV.
These three ranges are listed together with their respective detection
significances of the Crab nebula as calculated with Eq.~$17$ of
ref.~\autocite{1983_LiMa} and the respective angular resolutions in
Supplementary Table~2.

For the subsequent morphology fit, two maps are produced: One
containing all gamma-ray candidates (ON map), and one with the
gamma-ray-like background, estimated with an improved version of the
\textit{ring background} technique~\autocite{2007_Berge_Background},
which automatically adapts the ring size. The bin size of the maps is
$0.01^{\circ} \times 0.01^{\circ}$, well below the width of the point
spread function (PSF). We have verified that smaller bin sizes
have no influence on the subsequent results. For visualisation
purposes, the projected distribution of resulting events as a function
of squared angular distance ($\vartheta^2$) to the centroid of the
measured gamma-ray excess is also calculated. This distribution for
the $0.7 < E < 30\,\mathrm{TeV}$ energy range is shown in
Fig.~\ref{fig:ThetaSq}. The best-fit position in J2000 coordinates is
$\alpha = 5\mathrm{h}34\mathrm{m}30.9\mathrm{s} \pm \left(
  0.1\mathrm{s} \right)_{\mathrm{stat}} \pm \left( 1.3\mathrm{s}
\right)_{\mathrm{sys}}$,
$\delta = + 22^{\circ}00'44.5'' \pm 1.1''_{\mathrm{stat}} \pm
20''_{\mathrm{sys}}$ (systematic error from ref.~\autocite{2004_Gillesen}),
which is within uncertainties compatible with the Crab pulsar
location.

With dedicated Monte-Carlo (MC) simulations of the data-set, including
the actual instrument and observation conditions at the time of the
observations and using a power-law energy
distribution~\autocite{2017_Holler_RWS}, we re-weight the simulated events
to mimic the shape of the Crab nebula's energy spectrum and analyse
them with the same algorithms and analysis configurations as the
actual data. The resulting $\vartheta^2$ histogram of this MC analysis
serves as the PSF for this source and data-set and is also shown in
the upper panel of Fig.~\ref{fig:ThetaSq}. The $68\%$ and $90\%$
containment radii of our PSF are given in Supplementary Table~2.

As apparent in Fig.~\ref{fig:ThetaSq}, the PSF is highly inconsistent
with the distribution of the gamma-ray excess counts. The residuals in
the lower panel indicate clearly that the data are shallower than the
PSF. To study this further, we perform a two-dimensional morphology
fit with \textit{Sherpa}~\autocite{2001_Sherpa}, using the ON map, the
background map, and the simulated PSF. The PSF is convolved with a
two-dimensional radially symmetric Gaussian:
\begin{equation}
  \frac{\mathrm{d}P}{\mathrm{d}\vartheta^2} =
  \frac{1}{2\sigma_{2\mathrm{D,G}}^2}\cdot\exp\left( -
    \frac{\vartheta^2}{2\sigma_{2\mathrm{D,G}}^2}  \right)\, . 
\end{equation} 

To quantify the compatibility of the data and the convolved PSF, a
likelihood value is calculated and minimised. The best-fit extension
is found to be
$\sigma_{2\mathrm{D,G}} = 52.2'' \pm 2.9''_{\mathrm{stat}} \pm
6.6''_{\mathrm{sys}}$, with a preference of an extension of the Crab
nebula over a point-source assumption of $\mathrm{TS} \approx 83$. As
systematic uncertainty of the extension we quote the quadratic sum of
uncertainties related to the calibration and analysis method, to the
spectral shape used to re-weight the MC PSF, and to the fit method.

The resulting best-fit convolution is also plotted in
Fig.~\ref{fig:ThetaSq}. It clearly provides a good description of the
data both in the upper panel and the residuals in the lower panel. 

To verify the robustness of our result, we applied the analysis using
time-dependent simulations to two other bright and highly significant
extragalactic point-like gamma-ray sources, the active galactic nuclei
PKS 2155-304 and Markarian~421.  As illustrated in Supplementary
Figure~1, both sources appear to be point-like, while the Crab PWN
data is very clearly extended.  Upper limits on the extension of PKS
2155-304 and Markarian~421 are shown in Supplementary Figure~2. These
are well below the measured extension of the Crab nebula. We emphasise
that Markarian~421 culminates at large zenith angles of
$\phi > 60^{\circ}$ at the H.E.S.S. site (as opposed to
$\phi \approx 47^{\circ}$ for the Crab nebula culmination), making
this source a particularly convincing test of our PSF understanding,
since larger zenith angle observations have larger systematic PSF
uncertainties. As we also show in Supplementary Figure~2, we tested
the Crab nebula data-set for a zenith-angle dependence by splitting
the observations in two data-sets above and below $46^{\circ}$. The
measured extensions are compatible with each other.

The results have also been cross-checked with an independent calibration,
reconstruction, and analysis method~\autocite{2014_ImPACT}. We find
this second extension measurement slightly larger than our nominal
value (see Supplementary Figure~2), and use this difference as an
estimate of the systematic uncertainty related to the analysis method.

\textbf{Data and Code Availability Statement:} The raw data and the
code used in this study are not public but belong to the \hess\
collaboration. All derived higher level data that are shown in plots
will be made available on the \hess\ collaboration's web site upon
publication of this study.

\printbibliography[resetnumbers=false,heading=subbibliography,title={Additional references}]

\end{refsection}

\clearpage

\begin{refsection}

\section*{Supplementary Information}
\subsection*{Analysis and Results}
\begin{figure*}[b]
  \centering
  \includegraphics[width=0.9\textwidth]{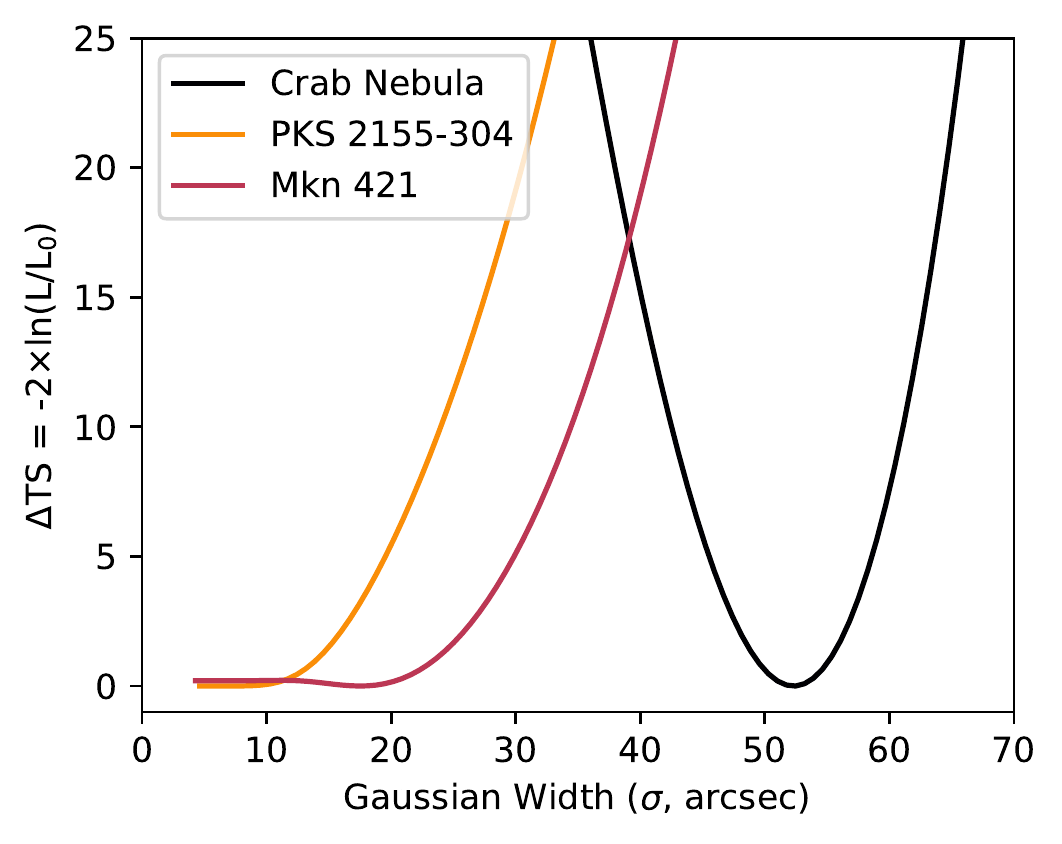}
  \caption{\textbf{Results of the extension fits.} Shown are the
    extension profiles of the difference 
    in test statistics ($\Delta TS = -2\times \ln(L/L_0)$) for the
    three sources evaluated here. The likelihood $L_0$ corresponds to
    the value at the respective minimum.}
  \label{fig:TS}
\end{figure*}

\begin{figure*}
  \centering
  \includegraphics[width=0.9\textwidth]{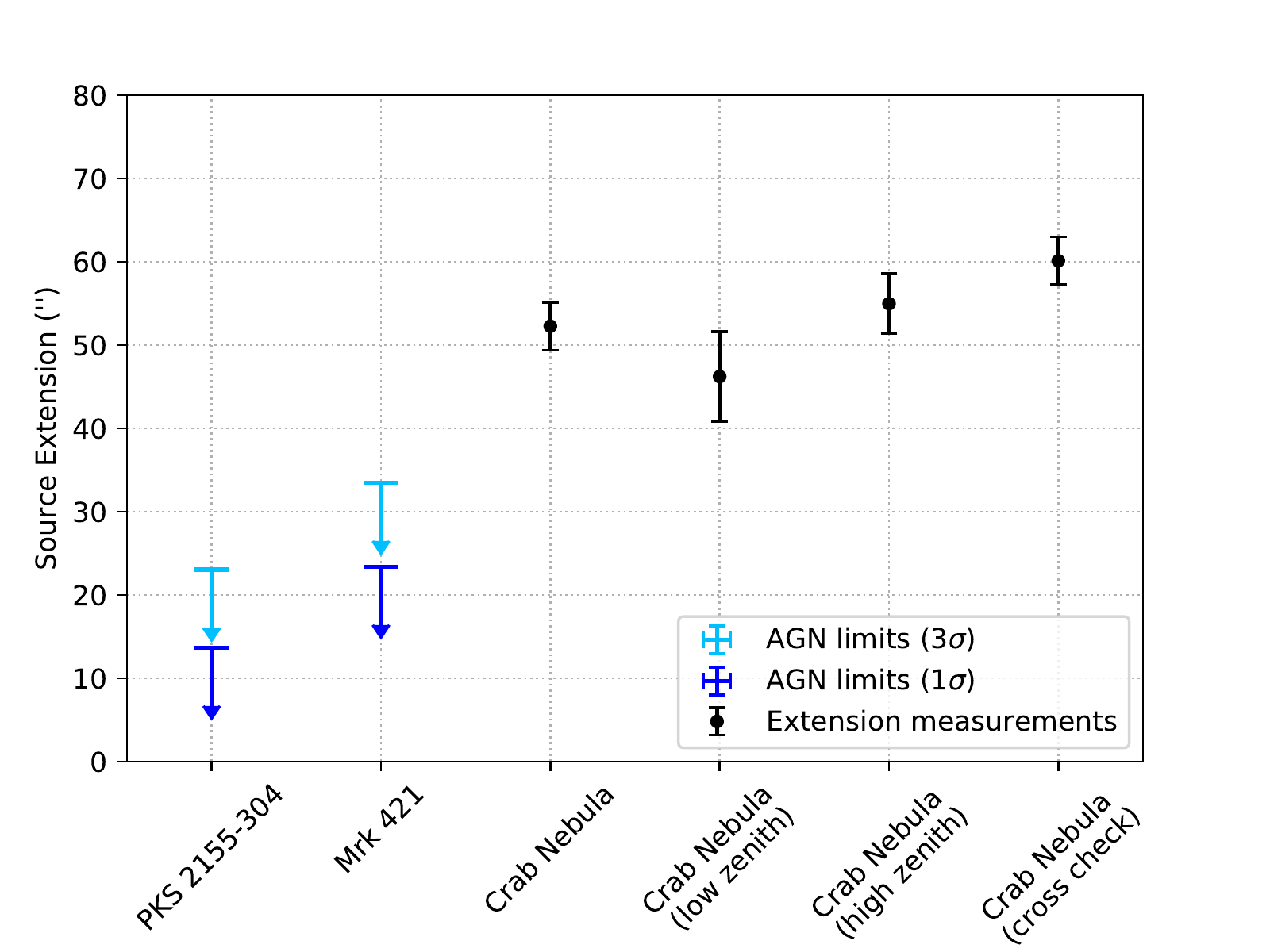}
  \caption{\textbf{Systematic checks of the extension measurement.}
    We present the derived extension upper limits of
    PKS~2155-304 and Markarian~421. The blue and cyan symbols
    are the 1 and 3 standard deviations upper limits, respectively.
    Shown in addition are the measured extension of  the Crab nebula
    and systematic checks. The low and high zenith angle band
    correspond to $44-46^{\circ}$ and $46-55^{\circ}$, respectively.
    The error bars are statistical uncertainties (1 standard
    deviation).}
  \label{fig:pointlike}
\end{figure*}

Figures~\ref{fig:TS} and \ref{fig:pointlike} demonstrate the
robustness of our Crab nebula extension measurement. In
Fig.~\ref{fig:TS} we show that we can clearly separate extragalactic
point-like gamma-ray sources like the active galactic nuclei
PKS 2155-304 and Markarian~421 from the extended Crab pulsar wind
nebula. In Fig.~\ref{fig:pointlike} we show in addition that the
measured extension is robust under variations of observation
conditions and analysis chains.

\noindent Tables~\ref{table:obstimes} and \ref{table:ebands} detail the dataset
and gamma-ray energy regimes investigated in this study.
\clearpage
\begin{table}[btp]
  \begin{tabularx}{\textwidth}{@{\extracolsep{\fill}}lccc} \toprule
    \multicolumn{1}{l}{year} &
                               \multicolumn{1}{l}{mean offset} &
                                                                 \multicolumn{1}{l}{mean zenith angle} &
                                                                                                         \multicolumn{1}{l}{livetime} \\
    \multicolumn{1}{l}{} &
                           \multicolumn{1}{c}{(degrees)} &
                                                           \multicolumn{1}{c}{(degrees)} &
                                                                                           \multicolumn{1}{c}{(hours)} \\\midrule
    2004 & 0.5 & 47 & 11.3 \\
    2007 & 0.5 & 46 & 3.6 \\
    2008 & 0.5 & 48 & 1.3 \\
    2009 & 0.5 & 46 & 7.1 \\
    \textbf{all} & 0.5 & 47 & 22.3 \\ \bottomrule
  \end{tabularx}
  \caption[]{\textbf{Overview of the \hess\ observation
    campaigns used in this study.} The livetime 
    given in hours corresponds to the data fulfilling data quality
    requirements.} 
  \label{table:obstimes}
\end{table}
\begin{table}
  \begin{tabular*}{\textwidth}{@{\extracolsep{\fill}}lccc} \toprule
    Energy Range & Detection Significance & $R_{68}$ (${}^{\circ}$) & $R_{90}$ (${}^{\circ}$) \\ \midrule
    $0.7\,\mathrm{TeV} < E < 30\,\mathrm{TeV}$  & $136.8\sigma$  & $0.052$ & $0.088$  \\
    $0.7\,\mathrm{TeV} < E < 5\,\mathrm{TeV}$ & $131.4\sigma$ & $0.053$ & $0.088$ \\
    $5\,\mathrm{TeV} < E < 30\,\mathrm{TeV}$ & $39.2\sigma$ & $0.043$ & $0.081$ \\  \bottomrule
  \end{tabular*}
  \caption{\textbf{Definition of the energy bands used for the analysis.} The
    detection significances of the Crab nebula and the angular
    resolutions expressed by the 68\% and 90\% containment radii
    ($R_{68}$ and $R_{90}$) of the simulated PSF are also given.}
  \label{table:ebands}
\end{table}

\clearpage

\subsection*{On the PWN Radiation Modelling}
\subsubsection*{Introduction}
Modern models for PWN have been developed over
the last forty years. They are based on the concept of a pulsar wind,
an ultrarelativistic outflow that connects the pulsar magnetosphere to
an outer nebula, which can be up to a few parsec in
size~\autocite{1974MNRAS.167....1R}, and on an MHD treatment of the
PWN~(described in ref.~\autocite{1984ApJ...283..710K}, called
\emph{KC2} in the following). In this model, the transport and
radiative cooling of high-energy particles is consistently described
with the analytical solution for the underlying MHD flow. The KC2
model for the non-thermal particles in the nebula allowed computing
the volume emissivity of synchrotron radiation. The emissivity
appeared to be quite sensitive to the properties of the MHD flow, in
particular to its magnetisation. The spectra predicted by the model
agreed well with observations provided that the pulsar wind is weakly
magnetised and ultra-relativistic. Later,
ref.~\autocite{1996MNRAS.278..525A} extended the approach of KC2 and
computed self-consistently the IC emission of the high-energy
particles. The spectra obtained agreed with observations in a vast
range, from optical wavelengths to the very high energy gamma-ray
band. This success gave strong support to MHD models and made a strong
case for efficient acceleration of charged particles to very high
energies by relativistic shock in PWNe.

The model of KC2 contains, however, an obvious shortcoming. It
utilizes an internally inconsistent model since one-dimensional MHD
models cannot include a toroidal magnetic field. This contradiction
can be resolved with two- or three-dimensional MHD models. Moreover,
there is another argument for a multi-dimensional MHD description. The
energy flux in the pulsar wind should be highly anisotropic with the
most significant fraction of energy released into a relatively small
range of solid angles close to the equatorial
plane~\autocite{2002MNRAS.336L..53B,2002MNRAS.329L..34L}.
Ref.~\autocite{2002MNRAS.336L..53B} suggested that a simple MHD model
that utilizes the analytical solution of KC2, limited to a region
close to the equatorial plane, can qualitatively reproduce the bright
torus seen in the X-ray energy band with
Chandra~\autocite{2000ApJ...536L..81W}. The formation of the jet-like
plumes seen in these Chandra data of the Crab PWN is then likely
caused by magnetic
collimation~\autocite{2002MNRAS.329L..34L,2003AstL...29..495K}.

Further on, the model of KC2 was extended by numerical MHD
calculations in
two~\autocite{2004AA...421.1063D,2004MNRAS.349..779K,2005MNRAS.358..705B}
and three~\autocite{2014MNRAS.438..278P} dimensions. Although the
quantitative comparison of the three-dimensional numerical model with
observational data has not yet been performed, many features of the
numerical solution seem to have a clear association with some observed
phenomena. In particular, X-ray wisps are robustly associated with MHD
waves propagating in the nebula.

To verify the potential of the TeV gamma-ray data, which reveal the
extension of the nebula, to constrain the allowed model parameter space, we
performed numerical simulations of energy spectra and the morphology
of the non-thermal electromagnetic emission. Such a study of a
multi-dimensional parameter space demands a computationally efficient
model. We have therefore adopted a one-dimensional MHD model as
developed in KC2. Since it is well known from the Chandra X-ray data
of the central part of the Crab nebula that the anisotropy of the
pulsar wind strongly influences the distribution of high-energy
particles, we introduced an additional parameter, \(\Delta \Omega\),
the solid angle into which the wind outflow propagates. Thus the wind
propagation region occupies a disk-like volume around the equatorial
plane. The region outside the disk is occupied by plasma that does not
yield any vital contribution to the X-ray emission and therefore we
ignore it.

We note that the one-dimensional numerical approach we take should be
considered as a {\it phenomenological model} in contrast to actual
{\it hypotheses} represented by more realistic three-dimensional MHD
simulations~\autocite{1980ConPh..21....3P}. Consequently, parameters
like the flow magnetisation parameter should be treated as internal
parameters of the phenomenological model that cannot be directly
compared to their values in two- or three-dimensional models. The flow
magnetisation for example, due to the rigid flow geometry, tends to
have smaller values in one-dimensional models, which are formally
inconsistent with the values revealed with more detailed
three-dimensional simulations. Applying a one-dimensional model is
then still worthwhile as it demonstrates the potential of the TeV
gamma-ray morphology data to constrain the model parameter and thus
verify the consistency of the tested model. The phenomenological model
used accurately accounts for processes governing the particle
emission. It allows us to link different radiation domains
(synchrotron, IC emission) and the energy-dependence of the emission
volumes visible in these domains. Thus, a hypothetical inconsistency
of the used phenomenological model with the X- and gamma-ray data
should be considered as a serious challenge for all MHD models for the
Crab Nebula. Finally, we note that when detailed three-dimensional
simulations of the synthetic emissivity in the nebula will be
available, an almost identical approach will help us to constrain the
allowed parameter space for these models.

\subsubsection*{MHD treatment}
In the framework of our model, the MHD flow in
the disk depends on three parameters. The first important parameter is
the wind magnetisation. This parameter determines the fraction of the
pulsar spin-down losses carried away in the form of a Poynting flux
\be
L\sd = \Delta\Omega n_1 \gamma_1 u_1 r\ts^2 m c^3 \left(1 + \sigma \right).
\label{eq:spindown_sigma}
\ee
Here \(L\sd\) is the pulsar spin-down (SD) losses, \(m\) is the
electron rest mass, and \(c\) is the velocity of light.  The
parameters with subscript \(1\) describe the flow upstream of the
termination shock (TS):  \(n_1\), \(\gamma_1\), and \(u_1\) are the
plasma density, the bulk Lorentz factor, and the four-velocity,
respectively. The wind magnetisation parameter is then:
\be
  \sigma=\frac{B_1^2}{4\pi n_1 \gamma_1 u_1 r\ts^2 m c^2}\,.
\ee

The TS radius, \(r\ts\), the wind opening angle, \(\Delta\Omega\), and
the wind magnetisation are the three parameters that determine the
model MHD solution. Chandra X-ray observations constrain the
\(r\ts\simeq0.1\)~--~\(0.15\rm \,pc\) at the pulsar wind equatorial
plane. Depending on the flow zenith angle, \(\theta\), the distance
between the pulsar and the TS can change considerably. However, since
the bulk of the emission is produced close to the equatorial plane, we
assume that the Chandra measurements define the physical range for the
model parameter \(r\ts\).

The wind opening angle is related to an anisotropy of the energy flux
in the pulsar wind, but other factors may also have a considerable
impact on it. For example, in the framework of more realistic two- or
three-dimensional MHD models, the shocked pulsar wind can be
significantly deflected towards the equatorial plane. This effect
cannot be consistently accounted for by the one-dimensional model used
here. Instead, we allow the model parameter \(\Delta \Omega\) to also
take smaller values than anticipated by the expected energy
anisotropy, which is expected to be proportional to \(\sin^2\theta\).

Downstream of the TS, approximating the magnetic field as toroidal,
the flow dynamics is described by the following system of
equations~\autocite{1984ApJ...283..710K}, which describes conservation
of particle flux, magnetic flux, adiabatic assumption, and total
energy, respectively:
\begin{align}
\dif[t]{\,} \left(cnur^2\right) &= 0, \label{eq:cons_num} \\
\dif[r]{\,} \left(\frac{ruB}{\gamma}\right) &= 0, \label{eq:cons_magnetic}\\
\dif[r]{\,} \left(nur^2e\right) + P\dif[r]{\,} \left(r^2 u\right) &= 0, \label{eq:prop_internal}\\
u\dif[r]{\,} \left(\gamma \epsilon\right) = \dif[r]{\,}
  \left[nur^2\left( \gamma \mu + \frac{B^2}{4\pi n\gamma} \right)
  \right] & = 0. \label{eq:cons_totalE} 
\end{align}
Here \(e\) is the specific internal energy per particle, $\mu$ is the
specific enthalpy ($\mu = \epsilon + p$), and $\epsilon$ is the sum of
the specific electromagnetic and internal energy in the proper frame. 
\(P\) is then the gas pressure and \(p\) the specific pressure.

KC2 have shown that the combination of the four equations given above
(\ref{eq:cons_num})~--~(\ref{eq:cons_totalE}) leads to the following
expression:
\be
\left(1 + u_2^2 v^2\right)^{1/2} \left[ \delta + \Delta \left(v
    z^2\right)^{-1/3} + \frac{1}{v} \right] = \gamma_2 \left(1 +
  \delta + \Delta\right)\,,\label{eq:algeb_eq} 
\ee
which determines the downstream flow velocity $u(z) = u_2 v$ as a
function of the dimensionless distance: $z = r/r\ts$. The subscript
\(2\) marks again the flow parameters downstream of the TS. The up-
and downstream parameters are related through the Rankine-–Hugoniot
conditions. The dimensionless parameters $\delta$ and $\Delta$ are
defined as
\begin{align}
\delta &= \frac{4\pi n_2 \gamma_2^2 mc^2}{B_2^2} \approx
         \frac{u_2}{u_1\sigma}\approx0, \\ 
\Delta &\equiv \frac{16\pi P_2 \gamma_2^2}{B_2^2} = \left(
         \frac{1+\sigma}{\sigma}\right) \frac{u_2}{\gamma_2} -1. 
\end{align}
Since the pulsar wind is expected to be ultra-relativistic,
\(\gamma_1>u_1\gg1\), and since the downstream velocity is determined
by the Rankine-–Hugoniot conditions, the equation~\eqref{eq:algeb_eq}
depends effectively only on the magnetisation $\sigma$. The two other
parameters that determine the normalisation factors are $r\ts$ and
\(\Delta \Omega\), the characteristic length scale and the geometric
extension of the emitting volume (that is, the flow).

\subsubsection*{Non-thermal particles}
We assume that particles up to and
beyond TeV energies in the Crab Nebula are accelerated at the pulsar
wind TS. The acceleration process results in a fixed distribution of
particles in the immediate vicinity of the TS. According to
ref.~\autocite{1996MNRAS.278..525A}, the spectral energy distribution
of the Crab Nebula is well reproduced by a parent electron
distribution following a broken power-law with exponential cutoff:
\be
n_0=\left.\dif[\ve]{N}\right|_{r=r\ts}=\left\{{ A\ve^ {-p\inj}
    \exp \left( -\frac{\ve}{\ve\cut} \right), \, \ve>\ve_{\rm b} \atop
    A(\ve/\ve_{\rm b})^{-\nicefrac32}\ve_{\rm b}^{-p\inj}, \, \ve<\ve_{\rm
      b}}\right.,
\ee
where \(\ve\) is the electron energy, \(A\) is a normalisation
constant, and \(\ve\cut\) is the cutoff energy.  Following
ref.~\autocite{1996MNRAS.278..525A}, we adopted
\(\ve\cut = 2.5 \times 10^{15}\,\rm eV\) and \(p\inj = 2.4\). At the
TS, the distribution normalisation, \(A\), and the break energy,
\(\ve_{\rm b}\), are adjusted so that the total number of particles
and internal energy of the non-thermal distribution equal the
values dictated by the Rankine-–Hugoniot conditions.

The electron energy distribution in the flow changes with distance
from the termination shock due to particle energy losses.  We consider
a differential volume element $\dif{V}$ at distance \(r\) from the TS:
\be
n(r, \ve) = n(r =r\ts, \ve_0) \dif[\ve]{\ve_0} \dif[V]{V_0} = n_0 \varphi \dif[\ve]{\ve_0},
\label{eq:eed}
\ee
where \(n_0=n(r =r\ts, \ve_0)\) is the initial electron energy
distribution, and $\varphi = \rho/\rho_0$ is the plasma compression
(that is, a parameter determined with MHD simulations). The subscript $0$
indicates the particles at TS $r=r\ts$, which corresponds to the
moment when the fluid element passes the TS and non-thermal particles
are accelerated.

The time evolution of particle energy \(\ve\) is described by the cooling equation:
\be
\dif[r]{\ve} = \frac1v\dot{\ve}(r,\ve),
\ee
where $\dot{\ve}$ is the energy loss rate. In PWNe, synchrotron (SYN),
inverse Compton (IC), and adiabatic (AD) energy losses represent the
most important cooling channels: 
\be
\dot{\ve}(r,\ve) = \dot{\ve}\syn(r,\ve) + \dot{\ve}\ic(r,\ve)+ \dot{\ve}\ad(r,\ve).
\label{eq:energy_loss_rate}
\ee
Further computational simplification can be achieved if one adopts the
Thompson approximation for IC cooling. In this case, one obtains 
\begin{align}
&\dot{\ve}\syn + \dot{\ve}\ic = - a\ve^2, \\
&a = \frac{4}{3} \frac{\sigmat c}{(m c^2)^2} \left(w\ph + w_{\rm B}\right),
\label{eq:syn_ic_loss}
\end{align}
where $\sigmat$ is the Thomson cross section, and $w\ph$ and $w_{\rm
  B}$ are the energy densities of the target photons and the magnetic
field, respectively. The adiabatic loss rate is given by
\be
\dot{\ve}\ad = \frac{v}{3} \dif[r]{\ln \rho} \ve,
\label{eq:ad_loss}
\ee
where $\rho$ is the plasma density in the fluid
element. Equation~\eqref{eq:energy_loss_rate} can now be rewritten by
using equations~\eqref{eq:syn_ic_loss} and \eqref{eq:ad_loss} as
\be
v\dif[r]{\,} \left(\frac{\mathrm{\rho^{1/3}}}{\ve}\right) = \rho^{1/3} (r) a(r).
\ee
Solving this equation, one obtains
\be
n(r, \ve) = \varphi^{4/3} \left(\frac{\ve_0}{\ve}\right)^2 n_0\,,
\label{eq:eed_2}
\ee
where $\ve_0$ is the initial electron energy:
\be
\ve_0 = \ve \frac{\varphi^{1/3}}{1 - \ve \lambda \rho^{-1/3}}.
\ee
The parameter \(\lambda\) accounts for radiative and adiabatic cooling
and is defined as
\be
  \lambda = \int\limits^{r}_{r\ts} \rho^{1/3}(r') a(r')\frac{\dif{r'}}{v(r')}\,.
\ee

\subsubsection*{Non-thermal radiation}
We aim to compute the spatial
extension of the non-thermal emission in the nebula. Since the plasma
emissivity varies considerably through the outflow, the emission
specific intensity and the total specific emission should be computed
by integrating over the line-of-sight (LoS) or the volume occupied by
the outflow:
\be\label{eq:emission_intensity}
I_\nu=\dif[t,\Omega,\nu,S]{E}= \int\limits_{\rm LoS}^{} j_\nu 
\dif{\ell}=\int\limits_{\rm LoS}^{}\left(\frac{\nu}{\nu'}\right)^2 j_{\nu'}'  \dif{\ell} \,,
\ee
where the primed variables correspond to the co-moving frame of the
plasma. In this case, the photon emission frequencies are affected by
a Doppler boosting factor: 
\be
\nu= \Db\nu'\,,
\ee
where \(\Db=1/\left[\G(1-\bb\ort{r}\obs)\right]\), \(\bb\) is the flow
bulk velocity, and \(\ort{r}\obs\) is a unit vector pointing towards
the observer. The dependence of the Lorentz factor on the flow
direction implies that the computation of the emission should be
performed in 3D geometry accounting for plasma bulk velocity in each
point of the nebula.

In the case of synchrotron radiation it is more convenient to consider
the emissivity in the fluid co-moving frame. In this frame one can use
the energy distribution to compute the radiation:
\be\label{eq:mec_general_2}
j'_{\nu'} = \int \dif[t',\nu']{N'_{\textsc{sp}}} \frac{n'(t',\br',\ve')}{4\upi} \dif{\ve'}\,.
\ee
Here, \(\dif[t',\nu']{N'_{\textsc{sp}}}\) represents the standard
single-particle spectrum of synchrotron radiation.

In the case of IC radiation, it is more convenient to define the
photon target in the laboratory frame, and consequently to compute the
emission directly in this frame. One obtains the emission  as 
\be\label{eq:mec_general_1}
j_\nu = \int \dif[t,\nu]{N_{\textsc{sp}}} f(\br,p\ort{r}\obs)
p^2\dif{p}\,,
\ee
where \(p=\sqrt{\ve^2-m^2}\) is the particle momentum and \(f\) is the
Lorentz-invariant distribution function:
\be
\dif{N} = f(\br,\bp) \dif{^3\br,^3\bp}\,.
\ee
This function can be expressed through the energy distribution in the
plasma co-moving frame as 
\be\label{eq:distr_function}
 f(\br,p\ort{r}\obs) \simeq\frac{c^3\Db^2}{4\upi \ve^2} n'\Big(\br,\ve /\Db\Big)\,.
\ee

The single-particle spectrum of IC emission can be obtained through
the angle averaged differential
cross-section~\autocite{1981ApSS..79..321A} as 
\be \label{eq:ic_sp_spectrum}
\dif[t,\nu]{N_{\textsc{sp:ic}}} = c\int
n\ph\left<\dif[\nu]{\sigma_{\textsc{ic}}}\Big(\ve,\nu,\epsilon\ph\Big)\right>\dif{\epsilon\ph}\,, 
\ee
where the target photon density accounts for all important
contributions like the cosmic microwave background, the far- and
near-infrared, and a synchrotron-self-Compton contribution: 
\be
n\ph = n\ph{}_,{}\cmb+n\ph{}_,{}\fir+n\ph{}_,{}\nir+n\ph{}_,{}\ssc\,.
\ee

\printbibliography[resetnumbers=false,heading=subbibliography,title={Supplementary Information references}]

\end{refsection}

\end{document}